# Enhancement of superconducting transition temperature and exotic stoichiometries in Lu-S system under high pressure

Juefei Wu[1], Bangshuai Zhu[1], Chi Ding[2], Dexi Shao[3], Cuiying Pei[1], Qi Wang[1,4], Jian Sun[2*], Yanpeng Qi[1,4,5*]

1. School of Physical Science and Technology, ShanghaiTech University, Shanghai 201210, China
2. National Laboratory of Solid State Microstructures, School of Physics and Collaborative Innovation Center of Advanced Microstructures, Nanjing University, Nanjing 210093, China
3. School of Physics, Hangzhou Normal University, Hangzhou 31121, China
4. ShanghaiTech Laboratory for Topological Physics, ShanghaiTech University, Shanghai 201210, China
5. Shanghai Key Laboratory of High-resolution Electron Microscopy, ShanghaiTech University, Shanghai 201210, China

* Correspondence should be addressed to Y.P.Q. (qiyp@shanghaitech.edu.cn) or J.S. (jiansun@nju.edu.cn)

## Abstract

Binary metal sulfides are potential material family for exploring high $T_c$ superconductors under high pressure. In this work, we study the crystal structures, electronic structures and superconducting properties of the Lu-S system in the pressure range from 0 GPa to 200 GPa, combining crystal structure predictions with *ab-initio* calculations. We predict 14 new structures, encompassing 7 unidentified stoichiometries. Within the S-rich structures, the formation of S atom cages is beneficial for superconductivity, with the superconducting transition temperature 25.86 K and 25.30 K for $LuS_6$-$C2/m$ at 70 GPa and $LuS_6$-$R$-$3m$ at 90 GPa, respectively. With the Lu/(Lu+S) ratio increases, the Lu-$d$ electrons participate more in the electronic properties at the Fermi energy, resulting in the coexistence of superconductivity and topological non-triviality of $LuS_2$-$Cmca$, as well as the superconductivity of predicted Lu-rich compounds. Our calculation is helpful for understanding the exotic properties in transition metal sulfides system under high pressure, providing possibility in designing novel superconductors for future experimental and theoretical works.

# Introduction

The pursuit for novel superconductors has been an important topic in condensed matter physics. The successful synthesis of $H_3S$ under 155 GPa with the superconducting critical temperature $T_c$ about 203 K [1, 2] after theoretical predictions [3, 4] accelerate the research on hydrogen metallization by "chemical precompression" from other elements [5], such as the binary hydrides with clathrate structures $CaH_6$, $YH_9$, $LaH_{10}$, etc. [6-13]. Meanwhile, Ref. 14 proposed that one can also look for breakthroughs in new classes of compounds. Such as borides [15-19], lithium-rich compounds [20-23], sulfides *etc.*

The simple substance sulfur is superconducting under high pressure, with $T_c$ about 15 K at 100 GPa [24]. Besides, the $T_c$ has the potential to reach about 17 K above 157 GPa [25] or 17.3 K at 200 GPa [26] by ac magnetic susceptibility measurements. The $T_c$ value of sulfur is the highest among the nonmetal elements except for hydrogen [27], and the metallization condition is less confined than hydrogen, indicating promising prospect in novel structure designing. As concluded in Ref. 28, it is tough for the $T_c$ values of the binary metal sulfides to exceed 5 K at ambient condition, with the 2M-$WS_2$ reaches 8.8 K [29]. Nevertheless, the $T_c$ values of $MoS_2$ and 2H-$TaS_2$ are above 10 K under high pressure, which are 12 K at 200 GPa [30] and 16.4 K at 157.4 GPa [31], respectively. Zhang *et al.* find that the superconducting state of PbS can be around 12 K at 19 GPa, which lowers the synthesized pressure from 1 Mbar [28]. In the aspect of theoretical studies, Gonzalez *et al.* explore the Sn-S system and predict two superconducting structures, $T_c$ values are 9.74 K and 21.9 K for SnS-*Pm*-3*m* at 40GPa and $Sn_3S_4$-*I*-42*d* at 30 GPa, respectively [32]. Thereafter, Matsumoto *et al.* confirm the pressure induced superconductivity of SnS-*Pm*-3*m* [33]. Besides, Shao *et al.* propose the exotic compound $Al_3S_4$-*I*-43*d*, which has the $T_c$ value 20.9 K at 100 GPa [34].

The transition metal could provide chemical pressure and lower the metallization pressure of hydrogen, supporting the high $T_c$ values of binary hydrides like $YH_9$ and $LaH_{10}$. This inspires the predictions in transition metal sulfides under high pressure. The high pressure synthesized $Y_3S_4$-*I*-43*d* shows weak superconductivity (3.6 K) [35], while Chen *et al.* predict an unconventional stoichiometric $YS_3$-*Pm*-3, which is 18.5 K at 50 GPa [36]. Gao *et al.* predict superconducting S-rich lanthanum sulfides under high pressure, the S cages stacked $LaS_3$ and $LaS_5$ are superconductors with $T_c$ of 13.6 K at 100 GPa and 11 K at 120 GPa, respectively [37]. Comparing with Al-S and Sn-S systems, transition metal has potential to elevate the S atoms ratio in binary metal sulfides with exotic structures. Since the last lanthanide element lutetium has similarity with lanthanum, which has the same unoccupied 5$d$ orbitals accompanied with a full 4$f$ shell, we wonder whether Lu can further improve the S atoms ratio in metal sulfides and enrich the binary sulfides family. Thus, it could be intriguing to explore the novel structures and exotic properties of the Lu-S system under high pressure.

In this work, we focus on exploring the Lu-S binary metal sulfides within 200 GPa, combining the machine learning graph theory accelerated crystal structure search with first-principles calculations. We predicted 14 new structures and 7 novel stoichiometries in the Lu-S system. There are potential superconductors in $LuS_7$, $LuS_6$, $LuS_3$, $LuS_2$, $Lu_5S_3$, $Lu_2S$ and $Lu_3S$ under high pressure. The $T_c$ values of the meta-stable $LuS_6$-$C/2m$ and $LuS_6$-$R$-$3m$ are about 25 K under high pressure, and we find the coexistence of superconductivity and topological non-triviality in $LuS_2$-$Cmca$ under high pressure. Combining with the calculations of the electronic structures and electron-phonon coupling (EPC) properties, we find that Lu-$d$ electrons are more involved in the electronic properties with the increasing of the Lu/(Lu+S) ratio, which alters the mechanism of superconductivity from the S-rich compounds to Lu-rich compounds.

## Methods

We used the machine learning graph theory accelerated crystal structure search method (Magus) [38, 39] to perform the variable compositions structure searches for lutetium sulfides under 50, 100, 150 and 200GPa, respectively. Then we carried out structure searches for the specific compositions, with more than 1000 structures evolved in 30 generations. Within Magus runs, we combined the Vienna *Ab-initio* Simulation Package (VASP) based on the density functional theory [40, 41] to perform the structure relaxations. The exchange-correlation functional was treated by the generalized gradient approximation of Perdew, Burkey, and Ernzerhof [42], and the projector-augmented wave (PAW) approach [43] was used to describe the core electrons and their effects on valence orbitals. The cutoff energy of the plane-wave was set to 400 eV and the sampling grid spacing of the Brillouin zone was $2\pi \times 0.05$ Å$^{-1}$.

We recalculate the enthalpy of the predicted structures to establish the convex hull. The convergence criteria were enhanced with the plane-wave kinetic-energy cutoff set to 600 eV, and the Brillouin zone sampling resolution was of $2\pi \times 0.03$ Å$^{-1}$. The convergence tolerance was $10^{-6}$ eV for total energy and 0.003 eV/Å for all forces.

The electronic structure calculations used a denser $k$-mesh grid of $2\pi \times 0.02$ Å$^{-1}$ and the total energy was converged to be less than $10^{-8}$ eV. We construct tight-binding models based on maximally localized Wannier functions (MLWFs) using WANNIER90 code [44]. The topological electronic structures are studied by the WANNIERTOOLS package [45].

The phonon spectrum were calculated by the PHONOPY [46] program package using the finite displacement method. The supercell was $2 \times 2 \times 2$ for all the candidate structures. The EPC coefficients were calculated by the QUANTUM-ESPRESSO (QE) package [47] using density-functional perturbation theory [48]. We selected the ultrasoft pseudopotential with a kinetic energy cutoff of 60 Ry. To estimate the superconducting transition temperature $T_c$ of the predicted structures, we used the self-consistent solution of the Eliashberg equation [49] and Allen-Dynes modified McMillan formula [50].

$$T_c = \frac{\omega_{\log}}{1.2} \exp\left(\frac{-1.04(1+\lambda)}{\lambda - \mu^*(1+0.62\lambda)}\right) \quad (1)$$

where $\omega_{\log}$ is the logarithmic average frequency, and $\mu^*$ is the Coulomb pseudo-potential for which we used the widely accepted values 0.10 [36, 37]. The EPC constant $\lambda$ and $\omega_{\log}$ are defined as below.

$$\lambda = 2 \int_0^\infty \frac{\alpha^2 F(\omega)}{\omega} d\omega \quad (2)$$

and

$$\omega_{\log} = \exp\left(\frac{2}{\lambda} \int \frac{d\omega}{\omega} \alpha^2 F(\omega) \ln(\omega)\right) \quad (3)$$

## Results and Discussion

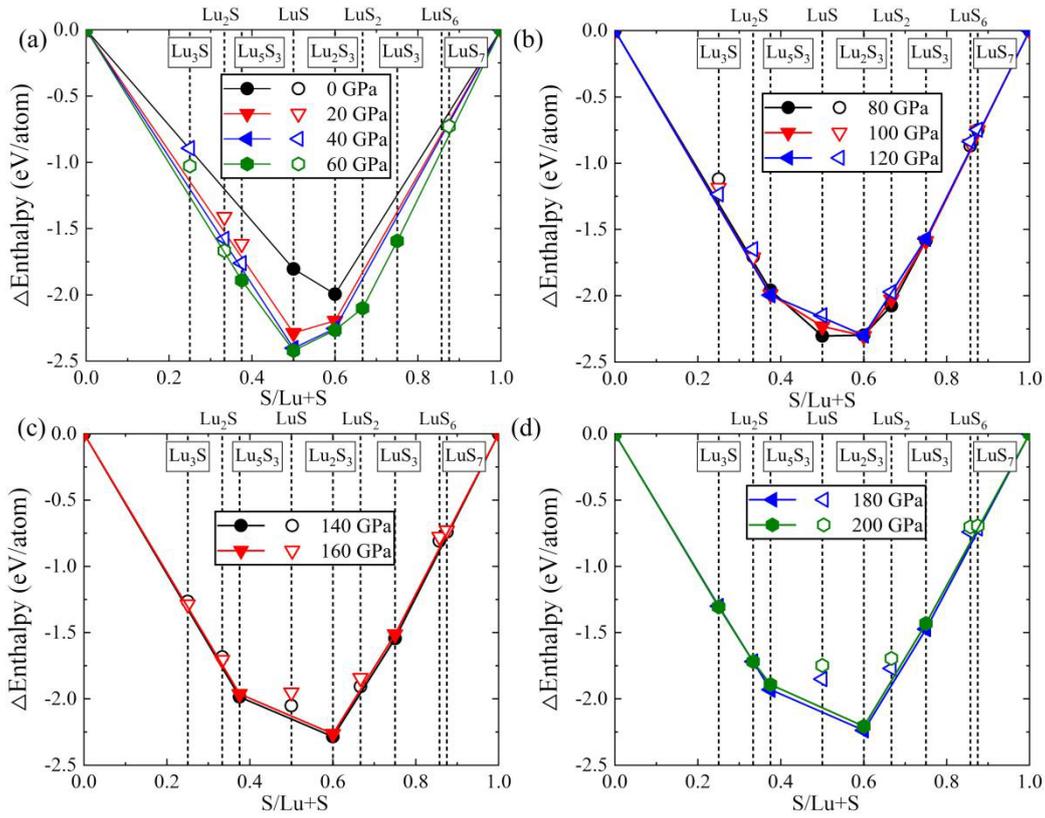

FIG. 1. The convex hull diagrams for Lu-S binary compounds under high pressures. The solid points indicate that the stoichiometries are on the convex hull, while the empty points are above the convex hull.

To investigate the stable phases in the Lu-S system under high pressure, we calculated the formation enthalpies of all the predicted structures by the variable composition structure searches under 50, 100, 150 and 200 GPa, and construct the convex hull diagrams in interval of 20 GPa. As shown in Fig. 1, we selected the simple substance Lu and S [51-55] under corresponding pressures as the references. At ambient pressure, only the known compounds in LuS and $Lu_2S_3$ are on the convex hull. Above 20 GPa, new stoichiometries emerge and they approach the convex hull with the increasing of pressure. The predicted stoichiometries $LuS_3$, $LuS_2$ and $Lu_5S_3$ lie on the convex hull at 60 GPa. Upon further compression, the predicted stoichiometries $Lu_2S$ and $Lu_3S$ are less than 50 meV/atom above the convex hull. $Lu_2S$ and $Lu_3S$ lie on the convex hull at 200 GPa and 180 GPa, respectively. Despite the predicted stoichiometries $LuS_6$ and $LuS_7$ is not on the convex hull within 200 GPa, $LuS_6$ is 35 meV/atom above the convex hull and $LuS_7$ is 42 meV/atom from the convex hull at 80 GPa, indicating their meta-stability under high pressure.

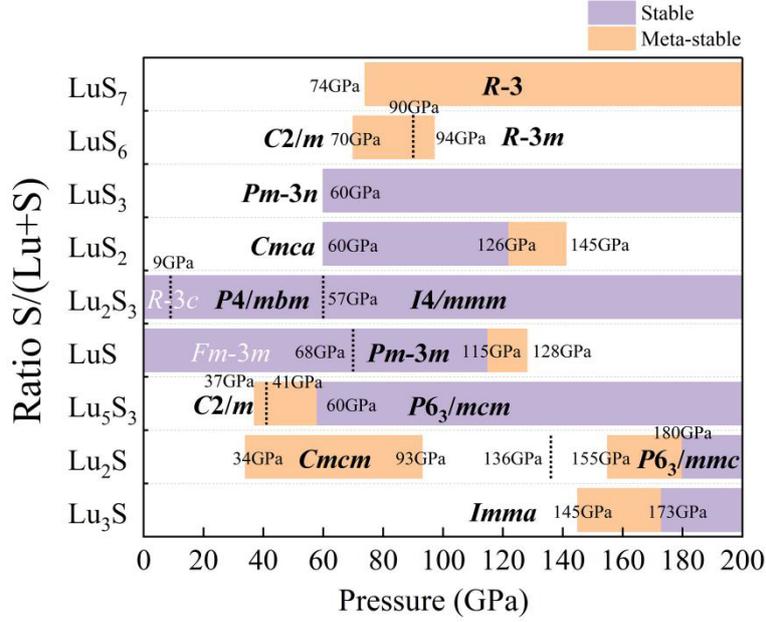

FIG. 2. The pressure-composition phase diagram of the Lu-S binary compounds. The predicted structures are in black bold font, while the reported structures are in white font. The violet region indicates the thermodynamically stable state and the orange region is the meta-stable state. The dashed lines are the boundary between different phases in each stoichiometry.

Combing the detailed calculations of relative enthalpy difference in each composition under high pressure [Fig. S1], we summarized the pressure composition phase diagram for Lu-S system in Fig. 2. We take the threshold of 50 meV/atom to estimate the meta-stable state, which is common in structure predictions [56]. Our calculations reproduce the stability of the reported structures LuS-$R$-3c and $Lu_2S_3$-$Fm$-3m at ambient condition and predicted 14 new structures $LuS_7$-$R$-3, $LuS_6$-$C2/m$, $LuS_6$-$R$-3m, $LuS_3$-$Pm$-3n, $LuS_3$-$Cmcm$, $LuS_2$-$Cmca$, $Lu_2S_3$-$P4/mbm$, $Lu_2S_3$-$I4/mmm$, LuS-$Pm$-3m, $Lu_5S_3$-$C2/m$, $Lu_5S_3$-$P6_3/mcm$, $Lu_2S$-$Cmcm$, $Lu_2S$-$P6_3/mmc$ and $Lu_3S$-$Imma$ within 9 stoichiometries [Fig. 2], the detailed lattice parameters and atomic coordinates are listed in Table S1. The phonon spectrum of the predicted structures are in Fig. S2-S5. The absence of imaginary frequencies in the whole Brillouin zone illustrates their dynamical stability within the calculated pressure range. Besides, we performed the spin-polarized calculations in all the predicted structures, none of them are magnetic within 200 GPa.

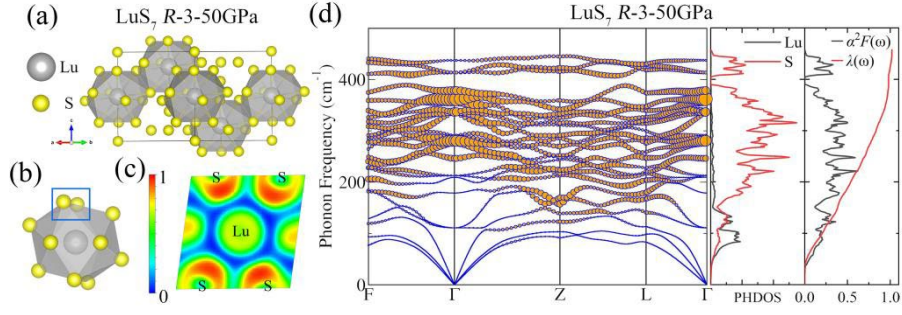

Fig. 3. (a) The crystal structure of LuS$_7$-$R$-3 at 50 GPa. (b) The S$_{12}$ cage of LuS$_7$-$R$-3. (c) The ELF of LuS$_7$-$R$-3. (d) The calculated phonon curves, PHDOS, Eliashberg spectral function $\alpha^2F(\omega)$, the electron-phonon integral $\lambda(\omega)$ of LuS$_7$-$R$-3 under 50 GPa. The size of the orange solid dots represents the contribution to electron phonon coupling.

As depicted in Fig. 2, the predicted LuS$_7$-$R$-3 structure is meta-stable from 74 GPa. Its relative enthalpy difference between the convex hull is decreasing with pressure, which is around 20 meV/atom at 200 GPa [Fig. S1 (a)]. The crystal structure of LuS$_7$-$R$-3 is shown in Fig. 3(a), S atoms surround the Lu atom forming S$_{12}$ cages [Fig. 3(b)]. The labeled S atom has a little shift and causes the distortion of the S$_{12}$ cages, which has discrepancy with the LaS$_{12}$ cages [37]. These cages stack with each other constituting the LuS$_7$-$R$-3 structure. The phonon spectrum indicates that LuS$_7$-$R$-3 structure can keep dynamical stability at 50 GPa [Fig. S2(a)]. We further calculated the band structures and the partial density of states (PDOS) at 50 GPa [Fig. S6(a)]. The $p$ electrons of S atoms make the main contribution in the range of -3 eV to 3 eV. The peaks in PDOS around Fermi energy may be beneficial for superconductivity. The electron localization function (ELF) results crossing the Lu-S$_4$ plane of the S$_{12}$ cages are in Fig. 3(c). There is almost no local charge between Lu and S atoms, suggesting ionic bonding properties, while the ELF between S atoms indicates covalent bonding and forms the S-S channel between S$_{12}$ cages. Then we carried out calculations on EPC constants, projected phonon density of states (PHDOS), Eliashberg spectral function $\alpha^2F(\omega)$ and the electron-phonon integral $\lambda(\omega)$ [Fig. 3(d)]. The vibration frequencies above 100 cm$^{-1}$ play dominant role in EPC, which corresponds to S atoms in the aspect of PHDOS. Hence the S$_{12}$ cages could be the key factor to the superconductivity in LuS$_7$-$R$-3. The calculated EPC constant $\lambda$ is 0.98 and the $T_c$ estimated by Allen-Dynes modified McMillan formula reaches 17.59 K [Table. 1].

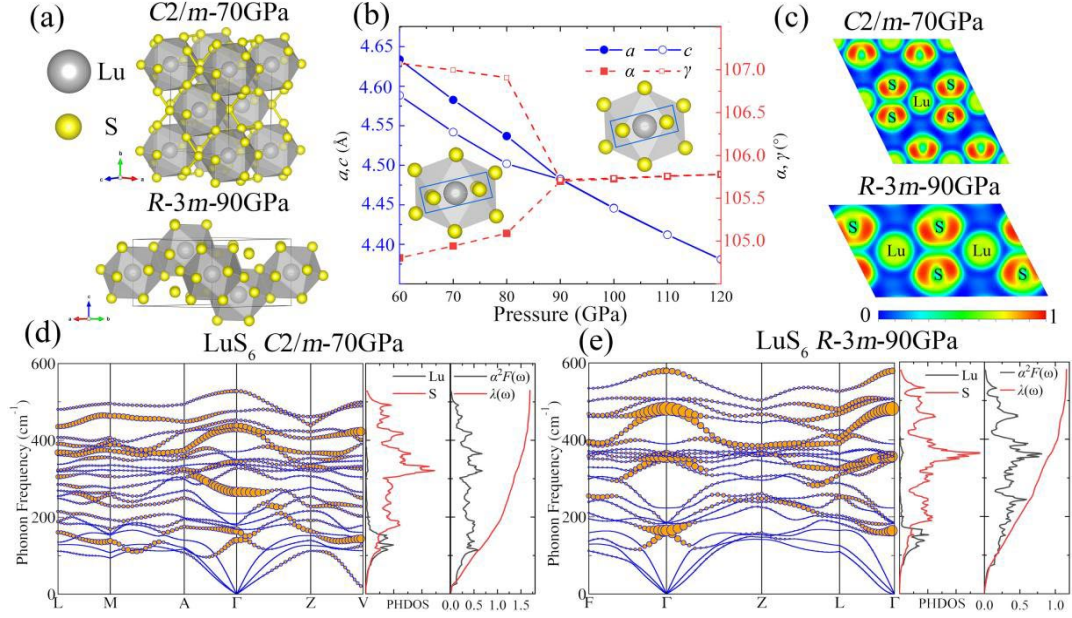

Fig. 4. (a) The crystal structure of LuS$_6$-$C2/m$ and LuS$_6$-$R$-$3m$ under high pressures. (b) The lattice parameters $a$, $c$, $\alpha$ and $\gamma$ of LuS$_6$-$C2/m$ in relation of pressure, the inset is the S$_{12}$ cage. (c) The ELF of LuS$_6$-$C2/m$ and LuS$_6$-$R$-$3m$ under 70 GPa and 90 GPa, respectively. (d) and (e) are the calculated phonon curves, PHDOS, Eliashberg spectral function $\alpha^2F(\omega)$, the electron-phonon integral $\lambda(\omega)$ of LuS$_6$-$C2/m$ at 70 GPa and LuS$_6$-$R$-$3m$ at 90 GPa, respectively.

We predicted two meta-stable structures LuS$_6$-$C2/m$ and LuS$_6$-$R$-$3m$ in LuS$_6$, both of them consist of S$_{12}$ cages as LuS$_7$-$R$-$3$. As plotted in the inset of Fig. S1(b), the relative enthalpy difference between LuS$_6$-$C2/m$ and LuS$_6$-$R$-$3m$ is less than 5 meV/atom above 60 GPa, which approaches our convergence limitation. Thus, we analyzed the lattice parameters $a$, $c$, $\alpha$ and $\gamma$ of LuS$_6$-$C2/m$ under high pressures [Fig. 4(b)]. After structure optimization, the curves of lattice parameters $a$ and $c$ merge with each other after 90 GPa and their difference is less than 0.1%, which is analogous between $\alpha$ and $\gamma$. We can observe a small alignment change of the S atoms in the S$_{12}$ cage after 90 GPa [inset of Fig. 4(b)], suggesting that LuS$_6$-$C2/m$ transforms to LuS$_6$-$R$-$3m$. In addition, the phonon spectrum calculations [Fig. S2(c)-(e)] illustrate the dynamical stability of LuS$_6$-$C2/m$ and LuS$_6$-$R$-$3m$ is from 70 GPa and 90 GPa, respectively, and we calculated their band structures, PDOS [Fig. S6 (b) and (c)] and ELF [Fig. 4(c)] under corresponding pressures. Analogous to LuS$_7$-$R$-$3$, the S-$p$ electrons make the main contribution around the Fermi energy, with the ionic bonding between Lu-S and covalent bonding between S-S. In the aspect of PHDOS, $\alpha^2F(\omega)$ and $\lambda(\omega)$ of LuS$_6$-$C2/m$ and LuS$_6$-$R$-$3m$ in Fig. 4(d) and (e), the dominant contribution to EPC is from S atoms above the 100 cm$^{-1}$ frequency region as well. This provides further evidence that S$_{12}$ cages are crucial to the superconductivity of the caged Lu-S structures. The calculated EPC constants $\lambda$ are 1.63 and 1.11, and the estimated $T_c$ are 25.86 K and 25.30 K for LuS$_6$-$C2/m$ at 70 GPa and LuS$_6$-$R$-$3m$ at 90 GPa [Table. 1], respectively. To our knowledge, this estimated $T_c$ values improve the prediction record in the binary transition metal sulfides [28, 35, 36, 57].

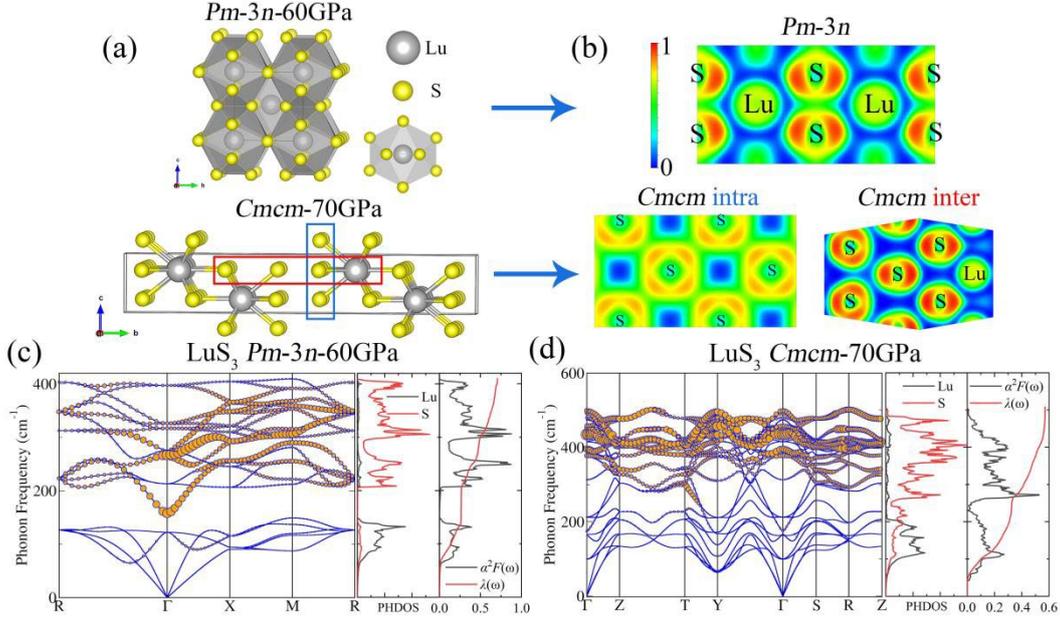

Fig. 5. (a) The crystal structure of LuS$_3$-$Pm$-$3n$ and LuS$_3$-$Cmcm$ under high pressures. (b) The ELF of LuS$_3$-$Pm$-$3n$, the inter and intra layer of LuS$_3$-$Cmcm$. (c) and (d) are the calculated phonon curves, PHDOS, Eliashberg spectral function $\alpha^2F(\omega)$, the electron-phonon integral $\lambda(\omega)$ of LuS$_3$-$Pm$-$3n$ at 60 GPa and LuS$_3$-$Cmcm$ at 70 GPa.

There are two predicted structures $Pm$-$3n$ and $Cmcm$ in LuS$_3$. Their relative enthalpy under high pressure are shown in Fig. S1(c). Despite LuS$_3$-$Cmcm$ has lower enthlapy than LuS$_3$-$Pm$-$3n$ below 71 GPa, the phonon spectrum calculations suggest the dynamical stability range of LuS$_3$-$Cmcm$ is above 70 GPa, which is above 60 GPa for LuS$_3$-$Pm$-$3n$ [Fig. S3(a)-(d)]. Hence LuS$_3$-$Cmcm$ is meta-stable and LuS$_3$-$Pm$-$3n$ is stable under high pressure. The crystal structures of LuS$_3$-$Pm$-$3n$ and LuS$_3$-$Cmcm$ are shown in Fig. 5(a). LuS$_3$-$Pm$-$3n$ composes of S$_{12}$ cages without distortion as in La-S system [37], and its ELF [Fig. 5(b)] has the typical S$_{12}$ cages characteristics, with Lu-S ionic bonding and S-S covalent bonding. LuS$_3$-$Cmcm$ is constructed by layers and the S atoms make up of the layer boundary. We calculated the ELF of intra and inter layer in LuS$_3$-$Cmcm$ [Fig. 5(b)]. The intra layer is S-S covalent bonding and forms channels, while connection type between inter layers is ionic. The band structures and PDOS of LuS$_3$-$Pm$-$3n$ and LuS$_3$-$Cmcm$ [Fig. S7 (a) and (b)] suggests the main role of S-$p$ electrons around the Fermi energy. Meanwhile, we can observe the distribution from the electrons of Lu atoms, which also has reflection in the EPC calculations. As shown in Fig. 5 (c) and (d), the dominant contribution to EPC constants is from the vibration above 200 cm$^{-1}$, which corresponding to S atoms in PHDOS, while the contribution to integral $\lambda(\omega)$ is larger than 40% in the Lu atoms region below 200 cm$^{-1}$, and it is particular in LuS$_3$-$Pm$-$3n$ [Fig. 5(c)]. This implies that the electrons of Lu atoms begin to have impact on the superconductivity with the increasing of Lu/(Lu+S) ratio. The EPC constants $\lambda$ are 0.69 and 0.52, with corresponding $T_c$ 9.57 K and 4.31 K for LuS$_3$-$Pm$-$3n$ at 60 GPa and LuS$_3$-$Cmcm$ at 70 GPa, respectively [Table. 1]. Comparing these two $T_c$ values, we assume that the caged structure is more favored for superconductivity than the layered structure in LuS$_3$.

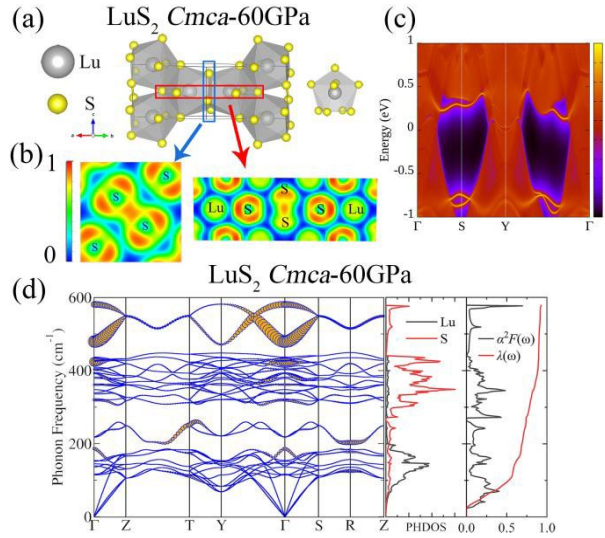

Fig. 6. (a) The crystal structure of LuS$_2$-$Cmca$ at 60 GPa. (b) The ELF of the S atoms plane and S-Lu-S plane. (c) The surface state of LuS$_2$-$Cmca$ at 60 GPa. (d) The calculated phonon curves, PHDOS, Eliashberg spectral function $\alpha^2F(\omega)$, the electron-phonon integral $\lambda(\omega)$ of LuS$_2$-$Cmca$ at 60 GPa.

The predicted structure LuS$_2$-$Cmca$ contains S$_9$ cages [Fig. 6(a)], which is isostructural to LaS$_2$-$Cmce$ [37]. LuS$_2$-$Cmca$ transforms to meta-stable state after 126 GPa [Fig. 1(d)] and is dynamically stable from 60 GPa [Fig. S3 (e) and (f)]. The ELF results in different planes suggest the channel between the S$_2$ dimer and weak bonding between Lu-S [Fig. 6(b)]. In particular, we can observe band nodes on the Fermi energy along the Brillouin paths $\Gamma$-$S$ and $T$-$Y$-$\Gamma$-$S$, and the bands contain different components S-$p$ and Lu-$d$, respectively. This is in line with PDOS results that the distribution of Lu-$d$ is comparable to S-$p$ around the Fermi energy, suggesting that electrons of Lu atoms participate more in the electronic properties in LuS$_2$ than the above predicted S-rich Lu-S compounds. Thereafter, we calculated the $\mathbb{Z}_2$ invariant with the spin-orbital coupling (SOC) for LuS$_2$-$Cmca$ at 60 GPa. The $\mathbb{Z}_2$ index ($\upsilon_0$;

$\upsilon_1\upsilon_2\upsilon_3$) = (1; 000), suggesting that LuS$_2$-*Cmca* is topologically non-trivial at 60 GPa. The surface states on (001) plane are in Fig. 6(c), we can observe surface states around *S* point and along the *Y-Γ* path. Moreover, we calculated EPC properties of LuS$_2$-*Cmca* at 60 GPa in Fig. 6(c), the Lu atoms contribute about 70% in the integral $\lambda(\omega)$. This is in agreement with PDOS results and Lu atoms is more involved in the electronic properties of LuS$_2$. The calculated $\lambda$ is 0.94 and the $T_c$ is 9.38 K at 60 GPa [Table. 1]. The coexistence of superconductivity and non-triviality in LuS$_2$-*Cmca* provides a potential platform for studying the relation between novel properties.

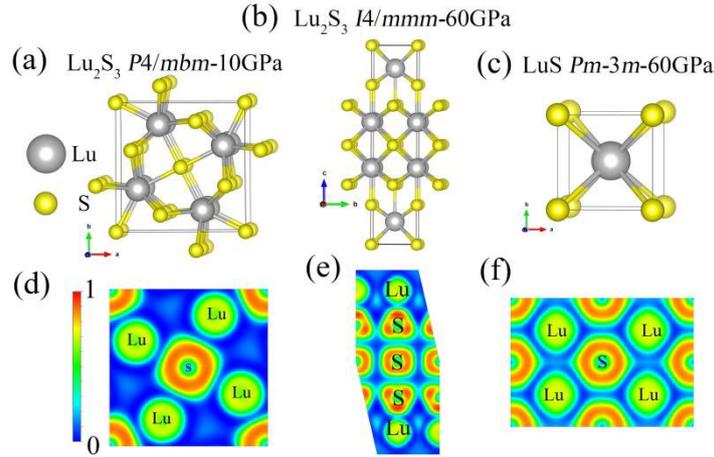

Fig. 7. (a)-(c) The crystal structures of Lu$_2$S$_3$-*P4/mbm*, Lu$_2$S$_3$-*I4/mmm* and LuS-*Pm-3m* under high pressures. (d)-(f) The ELF of the corresponding structures.

Lu$_2$S$_3$ and LuS are the two stoichiometries with ambient pressure structures [58, 59]. In Fig. S1 (e), the ambient phase Lu$_2$S$_3$-*R-3c* transforms to the predicted Lu$_2$S$_3$-*P4/mbm* around 9 GPa, then transforms to Lu$_2$S$_3$-*I4/mmm* after 57 GPa. Meanwhile, the ambient LuS-*Fm-3m* transforms to LuS-*Pm-3m* above 68 GPa, which becomes meta-stable about 115 GPa [Fig. S1(f)]. This is in line with the calculations in Ref. 60. Among the predicted structures in Lu$_2$S$_3$ and LuS [Fig. 7(a)-(c)], LuS-*Pm-3m* could keep dynamical stability at ambient pressure [Fig. S4]. According to ELF depicted in Fig. 7(d)-(f), the ionic bonding is predominant, indicating that Lu and S atoms are more isolated than other predicted S-rich compounds. The band structures and PDOS results of Lu$_2$S$_3$-*P4/mbm*, Lu$_2$S$_3$-*I4/mmm* and LuS-*Pm-3m* are in Fig. S8. In the predicted Lu$_2$S$_3$ structures, Lu-*d* and S-*p* have similar contribution around the Fermi energy. Lu-*d* and S-*p* are more distributed in conduction bands and valence bands, respectively. Considering the band crossings around the Fermi energy, we calculated the $\mathbb{Z}_2$ invariant and the index ($\upsilon_0$; $\upsilon_1\upsilon_2\upsilon_3$) = (0; 000) for Lu$_2$S$_3$-*P4/mbm* and Lu$_2$S$_3$-*I4/mmm*, illustrating the band structures are topologically trivial. Nevertheless, Lu-*d* becomes dominant in LuS-*Pm-3m* with a PDOS peak at Fermi energy, suggesting the growing impact of Lu-$d$ with the further increase of Lu/(Lu+S) ratio.

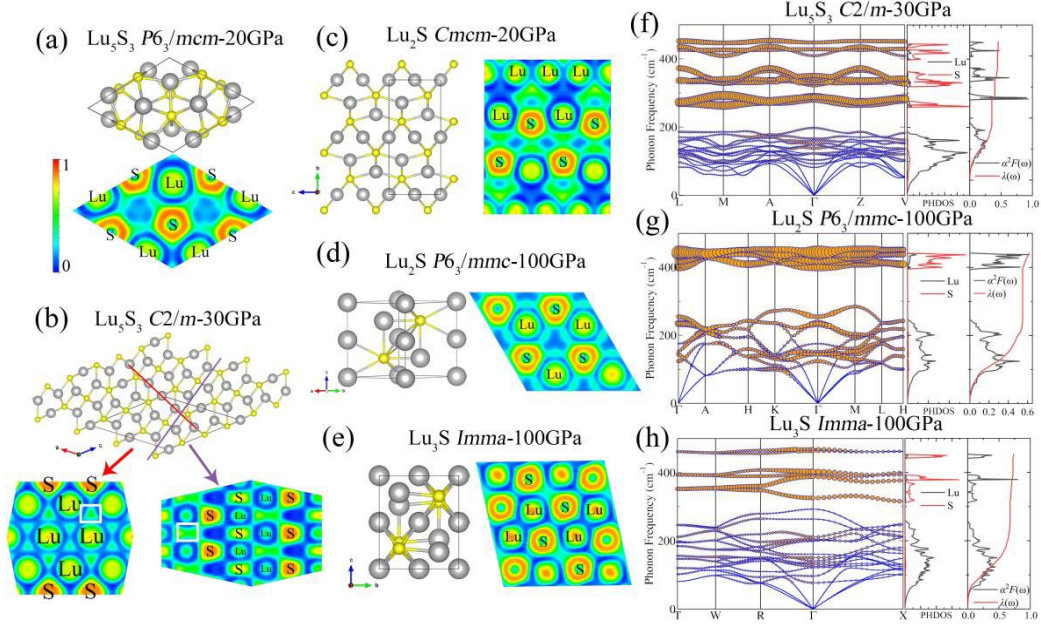

Fig. 8. (a)-(e) The crystal structure and corresponding ELF of the predicted structures Lu$_5$S$_3$-$C2/m$, Lu$_5$S$_3$-$P6_3/mcm$, Lu$_2$S-$Cmcm$, Lu$_2$S-$P6_3/mmc$ and Lu$_3$S-$Imma$ under high pressures. The ELF in (b) shows the channel connecting S atoms in different planes. (f)-(h) The calculated phonon curves, PHDOS, Eliashberg spectral function $\alpha^2F(\omega)$, the electron-phonon integral $\lambda(\omega)$ of Lu$_5$S$_3$-$C2/m$ at 30 GPa, Lu$_2$S-$P6_3/mmc$ at 100 GPa and Lu$_3$S-$Imma$ at 100 GPa.

As for the Lu-rich compounds in the Lu-S system, we predicted five new structures Lu$_5$S$_3$-$C2/m$, Lu$_5$S$_3$-$P6_3/mcm$, Lu$_2$S-$Cmcm$, Lu$_2$S-$P6_3/mmc$ and Lu$_3$S-$Imma$ in three novel stoichiometries [Fig. 8(a)-(e)]. The Lu$_5$S$_3$-$C2/m$ has lower enthalpy than Lu$_5$S$_3$-$P6_3/mcm$ above 41 GPa [Fig. S1(g)]. Although Lu$_5$S$_3$-$C2/m$ is 50 meV/atom above the convex hull below 37 GPa [Fig. 2], Lu$_5$S$_3$-$C2/m$ is dynamically stable from 30 GPa [Fig. S5(a)], and the pressure for Lu$_5$S$_3$-$P6_3/mcm$ is 20 GPa [Fig. S5(c)]. In Lu$_2$S [Fig. S1(h)], Lu$_2$S-$P6_3/mmc$ is more thermodynamically stable than Lu$_2$S-$Cmcm$ above 136 GPa, and the modification of the convex hull by Lu$_5$S$_3$-$P6_3/mcm$ separate the meta-stable region into two areas [Fig. 2]. Phonon spectrum results illustrate that the stabilized pressures start from 20 GPa and 100 GPa for Lu$_2$S-$Cmcm$ and Lu$_2$S-$P6_3/mmc$ [Fig. S5 (g) and (i)], respectively. Besides, the predicted Lu$_3$S-$Imma$ is dynamically stable after 100 GPa [Fig. S5(e)], while it enters meta-stable region above 145 GPa and becomes stable above 173 GPa [Fig. S1(i)]. The band structures and PDOS of Lu$_5$S$_3$-$C2/m$, Lu$_5$S$_3$-$P6_3/mcm$, Lu$_2$S-$Cmcm$, Lu$_2$S-$P6_3/mmc$ and Lu$_3$S-$Imma$ under high pressures are in Fig. S9. In all the five band structures, the Lu-$d$ electrons are predominant in the range of -3 eV to 3 eV. The valence bands and conduction bands overlap with each other around the Fermi energy, illustrating typical metal characteristics. Thus, we assume that the itinerant Lu-$d$ electrons give rise to the metallicity of Lu$_5$S$_3$-*C2/m*, Lu$_5$S$_3$-*P6$_3$/mcm*, Lu$_2$S-*Cmcm*, Lu$_2$S-*P6$_3$/mmc* and Lu$_3$S-*Imma*, which forms the connection between the S atoms [the ELF in Fig. 8(a)-(e)] and avoids the isolation in Lu$_2$S$_3$ and LuS. Moreover, we calculated the EPC properties of the five predicted structures. Among them, the predicted Lu$_5$S$_3$-*C2/m*, Lu$_2$S-*P6$_3$/mmc* and Lu$_3$S-*Imma* are superconducting under high pressure, as depicted in Fig. 8(f)-(h). Even though the vibration of S atoms has clear contribution to EPC, more than 80% contribution of the integral $\lambda(\omega)$ is from Lu atoms. This illustrates that the coupling between the S atoms and Lu-$d$ electrons is the elemental factor for superconductivity in the Lu-rich compounds, which is in agreement with the Lu-$d$ electrons induced metallicity. The EPC constants are 0.43, 0.61 and 0.73 for Lu$_5$S$_3$-*C2/m* at 30 GPa, Lu$_2$S-*P6$_3$/mmc* at 100 GPa, and Lu$_3$S-*Imma* at 100GPa, and the corresponding $T_c$ is 1.11 K, 4.78 K, and 6.91 K, respectively.

We summarized the calculated EPC constants and the estimated $T_c$ values of the superconducting Lu-S compounds in Table. 1. The $T_c$ values of the S-rich compounds are relatively higher than the Lu-rich compounds except for the layered LuS$_3$-*Cmcm*, indicating that the S cages are beneficial for the enhancement of superconductivity in Lu-S systems, such as the caged structures LuS$_7$-*R-3*, LuS$_6$-*C2/m*, LuS$_6$-*R-3m* and LuS$_3$-*Pm-3n*. With the increasing of Lu/(Lu+S) ratio, Lu-$d$ electrons involves more in the electronic properties in Lu-S compounds, resulting in the superconductivity and the band structures non-triviality of LuS$_2$-*Cmca*. The predicted structures in LuS and Lu$_2$S$_3$ are the transition zone, in which the Lu-$d$ and S-$p$ are comparable but the atoms are relatively isolated. In the Lu-rich compounds, the Lu-$d$ plays chief role in the electronic structures and the coupling between S atoms and Lu-$d$ electrons act as the key to the superconductivity, such as Lu$_5$S$_3$-*C2/m*, Lu$_2$S-*P6$_3$/mmc* and Lu$_3$S-*Imma*.

Table. 1. The calculated superconducting properties of the predicted Lu-S structures under high pressure with $\mu^*$ being 0.10.

| Phase | Space Group | Pressure (GPa) | $\lambda$ | $T_c$ (K) |
|---|---|---|---|---|
| LuS$_7$ | *R-3* | 50 | 0.98 | 17.59 |
| LuS$_6$ | *C2/m* | 70 | 1.63 | 25.86 |
| | *R-3m* | 90 | 1.11 | 25.30 |
| LuS$_3$ | *Cmcm* | 70 | 0.52 | 4.31 |

|  | Pm-3n | 60 | 0.69 | 9.57 |
| --- | --- | --- | --- | --- |
| LuS$_2$ | Cmca | 60 | 0.94 | 9.38 |
| Lu$_5$S$_3$ | C2/m | 30 | 0.43 | 1.11 |
| Lu$_2$S | P6$_3$/mmc | 100 | 0.61 | 4.78 |
| Lu$_3$S | Imma | 100 | 0.73 | 6.91 |

## Conclusion

In summary, we have explored the phase diagram of the Lu-S systems under high pressure combing with the machine learning graph theory accelerated structure searching and first-principles calculations. We predicted 14 novel structures, encompassing 7 exotic stoichiometries, all of which have the potential to be synthesized under high pressure. In the S-rich compounds, the S atoms formed cages are the key factor for the improvement of superconductivity, which is up to about 25 K for the meta-stable LuS$_6$-C2/m and LuS$_6$-R-3m, surpassing those of other binary metal sulfides. Lu-d electrons are more involved in the electronic properties in the predicted structures as the Lu/(Lu+S) ratio increases, and LuS$_2$-Cmca is both topologically non-trivial and superconducting. As for the Lu-rich compounds, the coupling between Lu-d electrons and S atoms become the crucial factor for superconductivity. Our work is helpful for understanding the transition metal sulfides under high pressure, providing some fundamental and novel insights for other metal sulfides in future experimental and theoretical works.

## Acknowledgement

This work was supported by the National Natural Science Foundation of China (Grant Nos. 52272265, U1932217, 11974246, 12004252), the National Key R&D Program of China (Grant No. 2018YFA0704300), and the Shanghai Science and Technology Plan (Grant No. 21DZ2260400).